\documentclass[prb,twocolumn,preprintnumbers,amsmath,amssymb]{revtex4-1}
\pdfoutput=1
\usepackage{epsfig,amsfonts}

\def\be{\begin{equation}}
\def\ee{\end{equation}}

\def\bea{\begin{eqnarray}}
\def\eea{\end{eqnarray}}

\def\ben{\begin{enumerate}}
\def\een{\end{enumerate}}

\def\bea{\begin{eqnarray}}
\def\eea{\end{eqnarray}}

\begin{document}

\title{Bethe Ansatz and Ordinary Differential Equation Correspondence for Degenerate Gaudin Models}
\author{Omar El Araby$^a$, Vladimir Gritsev$^a$, Alexandre Faribault$^b$}
\affiliation{$^a$ Physics Department, University of Fribourg, Chemin du Mus\'ee 3, 1700 Fribourg, Switzerland}
\affiliation{$^b$ Institute for Theory of Statistical Physics, RWTH Aachen, 52056 Aachen, Germany}
\affiliation{$^b$ Physics Department, ASC and CeNS, Ludwig-Maximilians-Universit\"at, 80333 M\"unchen, Germany}

\date{March 7, 2012}

\begin{abstract}
In this work, we generalize the numerical approach to Gaudin models developed earlier by us \cite{baode} to degenerate systems showing that their treatment is surprisingly convenient from a numerical point of view. In fact, high degeneracies not only reduce the number of relevant states in the Hilbert space by a non negligible fraction, they also allow to write the relevant equations in the form of sparse matrix equations. Moreover, we introduce a new inversion method based on a basis of barycentric polynomials which leads to a more stable and efficient root extraction which most importantly avoids the necessity of working with arbitrary precision. As an example we show the results of our procedure applied to the Richardson model on a square lattice.
\end{abstract}

\maketitle

\section{Introduction}
Gaudin-type models represent a fertile basis for an exact approach to the dynamics of quantum many-body systems. These models do not require fine-tuning of the Hamiltonian's parameters to satisfy integrability conditions. This is especially attractive from an applicational point of view since these parameters can represent physical coupling of the model. A particular example would be the central spin Hamiltonian\cite{centralspin} where the coupling constants between the central spin and the individual surrounding spins are related to the parameters of the Gaudin model in a simple way. We are talking about a non-degenerate model when all these couplings are different. Non-degenerate Gaudin-type models have received considerable attention recently. Apart from central spin physics they are relevant to a mesoscopic BCS\cite{exp}, Dicke model\cite{Dicke} , Lipkin-Meshkov-Glick and many other physical models\cite{ortiz}.  However in many physical situations there are natural degeneracies coming from the spectrum of the model. 
Thus, in the BCS model the bare dispersion relation $\epsilon_{d}({\bf{k}})$ of paired fermions on a d-dimensional finite lattice defines the coupling constants of the Gaudin model according to a simple equation, $\epsilon_{d}({\bf{k}})=\epsilon_{i}$. On a regular lattice this equation may have several solutions, and thus the spectrum of Gaudin parameters $\epsilon_{i}$ is degenerate. The degeneracy depends on the geometry of the lattice and is normally a number of the order of one. 

Recently we introduced a fruitful approach for solving Bethe equations for the Gaudin models \cite{baode} . The basic idea is to use an equivalent reformulation of the complex coupled Bethe ansatz equations in terms of an ordinary differential equation, a method known as BA/ODE correspondence \cite{dorey}. In the spirit of our previous work, here we generalize our approach to degenerate systems. It was previously shown \cite{faribault} that in certain non-equilibrium situations in the non-degenerate Richardson model the number of relevant eigenstates (those bringing a non-negligible contribution to the wave function) can be small drastically reducing the effective Hilbert space. However, in more general situations, a large number of eigenstates can be necessary therefore making an efficient solver desirable.

Degeneracies can play an important role in this respect since the number of solutions to the Bethe equations is automatically reduced compared to a equally large non-degenerate system. Moreover we show that the matrix equations involved become super-sparse for high degeneracies leading to further simplifications. This property allows us to improve the system size without loosing computational speed and study dynamics of extremely large systems compared to the usual Bethe ansatz solvable models's size for non-degenerate systems. 

In the end we are looking for polynomial solutions to a second order differential equation (see eq. (\ref{secondODE})) a procedure which is in every regard equivalent to the Heine-Stieltjes polynomial approach discussed in \cite{heine,heine2} for example. Here the differential equation depends on a non-trivial set of parameters which we first extract from the solutions of an ensemble of quadratic algebraic equations according to the procedure described in sections \ref{batoode} to \ref{proceed}. The first one summarizes the results of our previous paper \cite{baode}, while in the following we explicitly derive all the relevant equations for the degenerate case. The polynomial of interest can then be obtained using a new method based on the Lagrange polynomial basis which is explained in section \ref{rootex} and can be used for both degenerate or non-degenerate Gaudin model. The techniques discussed are finally applied to the Richardson model on a square lattice \ref{bcs}.

\section{From Bethe Ansatz to Ordinary Differential Equations}
\label{batoode}

This work deals with a set of quantum integrable models which fall in the rational family of Gaudin models defined by the following operator algebra
\begin{align}
[S^\kappa(\lambda_i),S^{\kappa+1}(\lambda_j)]&=ig\frac{S^{\kappa+2}(\lambda_i)-S^{\kappa+2}(\lambda_j)}{\lambda_i-\lambda_j}\nonumber\\
[S^\kappa(\lambda_i),S^\kappa(\lambda_j)]&=0,\kappa=x,y,z.
\label{algebra}
\end{align}
For any realization of the rational Gaudin algebra a set of $N$ commuting operators $R_i$ can be defined, therefore allowing the construction of exactly solvable Hamiltonians of the form $\sum_{i}\eta_iR_i$.

The reduced BCS Hamiltonian on a regular lattice
\begin{align*}
H=\sum_{\bf{k},\sigma}\epsilon_d(\bf{k})\hat{c}^\dagger_{\bf{k},\sigma}\hat{c}_{\bf{k},\sigma}-g\sum_{\bf{k},\bf{k}'}\hat{c}^\dagger_{\bf{k},\uparrow}\hat{c}^\dagger_{-\bf{k},\downarrow}\hat{c}_{-\bf{k}',\downarrow}\hat{c}_{\bf{k}',\uparrow}
\end{align*}
is a natural example of  a degenerate model built according to $(\ref{algebra})$ and will be illustrated in section \ref{bcs}. This paper's aim is to provide an efficient and numerically stable algorithm which allows one to find exact eigenstates of this particular type of models by exploiting its quantum integrability through the algebraic Bethe ansatz. 

In this formalism, by defining a pseudovacuum $\left|0 \right>$, one can write exact eigenstates of the system in a remarkably compact form defined by the repeated action of a Gaudin operator on 
\bea
\left|\left\{\lambda_1 ... \lambda_M\right\} \right>\propto \prod_i \mathrm{S}^+(\lambda_i) \left| 0 \right>.
\label{bethestate}
\eea
Here $\mathrm{S}^+(u)=S^x(u)+iS^{y}(u)$ plays the role of a generalized creation operator parametrized by a single complex variable $u$ whose explicit expression in terms of the fundamental operators defining a particular system will be model dependent \cite{baode,sierra,ortiz}. In every case, the pseudo vacuum $\left|0 \right>$ is defined as a lowest weight vector such that $\mathrm{S}^-(u) \left|0 \right> = 0$ for any value of $u$ which would be the Fock vacuum in the case of the reduced BCS Hamiltonian. This set of general states become eigenstates of the Hamiltonian provided the rapidities are solutions to the set of non-linear algebraic equations collectively known as Bethe equations. Specializing to rational Gaudin models, the general form of Bethe equations we address is given by
\begin{align}
F(\lambda_i)=\sum_{j\neq i}\frac{1}{\lambda_i-\lambda_j},
\label{bethe}
\end{align}
with 
\begin{align}
F(\lambda_k)=-\sum_{i=1}^N\frac{A_i}{(\epsilon_i-\lambda_k)}+\frac{B}{2g}\lambda_k+\frac{C}{2g}.
\label{F}
\end{align}

We will denote the ensemble of N distinct $\{\epsilon_j\}$ as "energy levels", $g$ as the "coupling constant" and $\mathcal{N}$ as the "total number of levels" (for non-degenerate systems $\mathcal{N}=N$, otherwise degeneracies make $\mathcal{N}>N$). While the physical nature of these parameters differs from model to model, in the Richardson case discussed here these parameters do indeed correspond respectively to single particle energy levels and coupling constant. 

To each distinct energy $\epsilon_j$ we can associate a degeneracy $d_j$ and a (pseudo-)spin magnitude $s_j$. The coefficients $A_j$ are then a product of those two values $A_j=s_j d_j$. In the remainder of this paper, we deal only with degenerate spin $1/2$ systems and we therefore fix all $s_j=1/2$ and have $A_j=d_j/2$. An arbitrary spin would simply modify $A_j$ in a way which is strictly equivalent to a larger degeneracy. One should note here that a degeneracy in the energy levels or a larger spin actually differ since in the degenerate case the solutions to the Bethe equations no longer form a complete basis of the Hilbert space. However the resulting subspace is orthogonal to the rest of the Hilbert space and therefore even for non-equilibrium problems, provided the initial condition only involves this subspace, its elements remain the only states required to study the unitary time evolution. This work only addresses the numerical issues associated with finding the particular subset of eigenstates corresponding to solutions of eq. (\ref{bethe}) applied to the particular pseudovacuum defined before.

Solutions of (\ref{bethe}) are found by scanning from the starting point $g=0$ for which the solutions are known ($\lambda_i\rightarrow\epsilon_k$ for the Richardson model) and going to finite $g$ via successive steps. Rapidities $\lambda_k$ can be real or pairs of complex conjugates, the points in $g$ at which two real rapidities form a complex pair (or vice versa) can lead to numerical instabilities when working directly with the $\lambda_k$ themselves. In \cite{dominguez} an algorithm which finds these points beforehand allowing one to control the equations in the critical regions has been proposed. Here however, in order to avoid these difficulties, we simply introduce the function 
\begin{align}
\Lambda(z)\equiv\sum_{k=1}^M\frac{1}{z-\lambda_k}=\frac{P'(z)}{P(z)}
\end{align}
where $P(z)=\prod_{k=1}^M(z-\lambda_k)$. The zeros of $P(z)$ then correspond to the roots $\lambda_k$ of the Bethe Equations (BA).

One can easily show that, whenever the set of rapidities ${\lambda_k}$ satisfies BA, the ODE
\begin{align}
\Lambda'(z)+\Lambda^2(z)-\sum_{\alpha=1}^N\frac{2F(\lambda_\alpha)}{z-\lambda_\alpha}=0
\label{riccati}
\end{align}
is satisfied.

If we restrict our investigation to non degenerate spin $1/2$ systems we would only need to solve the simple quadratic equations 
\begin{align}
\Lambda_j^2=g\sum_{i\neq j}\frac{\Lambda_j-\Lambda_i}{\epsilon_j-\epsilon_i}+\Lambda_j
\label{quadratic}
\end{align}
which are obtained by taking the limit $z\rightarrow\epsilon_j$ of Eq. (\ref{riccati}), setting $A_i=1/2~\forall i$ and $\Lambda_j=g \Lambda(\epsilon_j)$.

Once a set $\{\Lambda_j\}$, solution to Eq. (\ref{quadratic}) is found, the corresponding set of rapidities $\{\lambda_k\}$ can be obtained by solving only linear equations and running a standard root finding algorithm. While in previous publications\cite{baode,buccheri} it was suggested to use elementary symmetric polynomials to build the linear system, in section \ref{rootex} we propose to use a different polynomial basis which leads to much improved numerical stability.

\section{Degenerate systems}
\label{degen}

For the generalization to degenerate systems we obtain successive differential equations by taking the first $n^{th}$ derivatives of Eq. (\ref{riccati}), to do that let us fix some notations first. For each pair $(\epsilon_j,A_j)$ we then have to solve a system of equations which contains the first $d_j^{th}$ derivatives of (\ref{riccati}) and the first $d_j^{th}$ derivatives of $\Lambda_j$, this of course has to be done for all $j$.

The general form of the $n^{th}$ derivative of (\ref{riccati}) is derived in Appendix \ref{derivation}. Eq. (\ref{matej}) for $\Lambda^{(n)}_j=g^{n+1} \Lambda(\epsilon_j)^{(n)}\equiv g^{n+1} \left. \frac{d^{n}\Lambda(z)}{dz^{n}} \right|_{z=\epsilon_j}$ reads 
\begin{align}
\mathcal{E}^{(n)}_j(g)=& \kappa_j^{(n)}(g)+\sum_{k=0}^n\binom{n}{k}\Lambda_j^{(k)}\Lambda_j^{(n-k)}
\nonumber\\&-(\frac{Bn\epsilon_j}{g}\Lambda_j^{(n-1)}+C\Lambda_j^{(n)})
\nonumber\\&+(1-\frac{d_j}{n+1})\Lambda_j^{(n+1)} = 0,
\label{nthder}
\end{align}

with
\begin{align}
\kappa_j^{(n)}(g):=&-\sum_{i\neq j}^N d_i n! \left(g^{n+1}\frac{\Lambda_i-\Lambda_j}{(\epsilon_i-\epsilon_j)^{n+1}}\right.
\nonumber\\
&\left.-\sum_{k=1}^n g^k\frac{1}{(n+1-k)!}\frac{\Lambda_j^{(n+1-k)}}{(\epsilon_i-\epsilon_j)^k}\right).
\end{align}
  
The last term of (\ref{nthder}) cancels for $d_j=n+1$. Thus, given a set of pairs $\{(\epsilon_1,d_1),(\epsilon_2,d_2),\dots,(\epsilon_k,d_k)\}$, we define a closed set of quadratic algebraic equations by using, for each pair, the corresponding set of equations: i.e for $d_j=1$ we solve equation $\mathcal{E}^{(0)}$ which depends on $\Lambda_j^{(0)}$, for $d_j=2$ we solve $(\mathcal{E}^{(0)},\mathcal{E}^{(1)})$ in terms of $(\Lambda_j^{(0)},\Lambda_j^{(1)})$ and so on.

\section{Weak coupling limit of the BA equations}
\label{weak}

Once all the equations for $\Lambda_j$ and its derivatives are set, the fact that they are non-linear requires the use of an iterative method. To do so, one always needs an initial guess which constitutes a good enough approximation to guarantee convergence to the desired solution of the BA equations. As previously mentioned, this will be achieved by slowly increasing the coupling constant $g$ but to begin this process we need to solve the Bethe equations in the weak coupling limit for a general degenerate system.

The original form of BA equations is 
\bea
-\sum_{i\neq k}^M\frac{2}{\lambda_k-\lambda_i}+\sum_{i=1}^{N}\frac{d_i}{\lambda_k-\epsilon_i}+\frac{B}{g}\lambda_k+\frac{C}{g}=0
\label{ba}
\eea
where $N$ is the number of distinct energies $\epsilon_i$.

We know already that for $g=0$ the roots will simply correspond to the values of the single energy states $\lambda_k=\epsilon_{i_k}$ which are occupied (the (pseudo-)spins which are flipped with respect to the pseudo-vacuum). Linearizing the system for weak coupling by inserting the solution
\bea
\lambda_k=\epsilon_{j}+g\Delta_k
\label{ans}
\eea
\noindent in (\ref{ba}), multiplying the obtained equation by $g$ and taking the limit $g\rightarrow 0$,  the BA equations become
\bea
-\sum_{i\neq k}^{r}\frac{2}{\Delta_k-\Delta_i}+\frac{d_j}{\Delta_k}+B\epsilon_{j}+C=0
\label{prepol}
\eea 
where $r$ is the number of roots occupying the single energy $\epsilon_j$ at $g=0$.

It is now possible to introduce the polynomial $f(x)=\prod_{k=1}^r(x-\Delta_k)$ which has the property
\bea
\lim_{x\rightarrow\Delta_k}\frac{f''}{f'}=\sum_{i\neq k}\frac{2}{\Delta_k-\Delta_i}.
\eea

Thus equation (\ref{prepol}) can be written in the form of a differential equation (in the limit $x\rightarrow\Delta_k$)
\bea
d_jf'-x f''+(B\epsilon_j+C)x f'=0.
\eea
The function $F(x)=d_j f'(x)-x f''(x)+(B\epsilon_j+C)x f'(x)$ is a polynomial of the same degree $r$ as $f(x)$ and with the same roots, the two polynomial are thus proportional to each other and the coefficient of proportionality is $r$ (see \  \cite{altshuler}). Therefore we have $F(x)=rf(x)$
and the equation can be written as follows
\bea
xf''(x)-\left[d_j+(B\epsilon_j+C)x\right]f'(x)+rf(x)=0.
\label{fin}
\eea
The known solutions of equation (\ref{fin}) for $B=0$ and $C=1$ (which encompasses the central spin and Richardson models) are the generalized Laguerre polynomials $L^{-1-d_j}_r(x)$. Therefore, for weak coupling, the solutions to the BA equations correspond to $\lambda_k=\epsilon_j+g\Delta_k$ with $\Delta_k$ roots of the Laguerre polynomials $L^{-1-d_j}_r(x)$.

\section{How to proceed}
\label{proceed}
In order to solve the sets of quadratic equations defined by (\ref{nthder}) we use a combination of Taylor expansion to generate an approximative solution at $g+\delta g$ and Newton's method to refine this approximation. For both methods we need to solve a linear system of equations defined by the block-$N\times N-$matrix  

\begin{align}
S=\begin{pmatrix}
S_{11} & \cdots & S_{1N}\\
\vdots & \ddots & \vdots\\
S_{N1}& \cdots &S_{NN}
\end{pmatrix}
\end{align}
in which each pair of indexes $(i,j)$ defines a matrix 
\begin{align}
S_{ij}=
\begin{pmatrix}
\frac{\partial \mathcal{E}_i^{(0)}}{\partial\Lambda_j^{(0)}}&\cdots &\frac{\partial \mathcal{E}_i^{(0)}}{\partial\Lambda_j^{(d_j-1)}}\\
\vdots & \ddots &\vdots\\
\frac{\partial \mathcal{E}_i^{(d_i-1)}}{\partial\Lambda_j^{(0)}}&\cdots &\frac{\partial \mathcal{E}_i^{(d_i-1)}}{\partial\Lambda_j^{(d_j-1)}}\\
\end{pmatrix} 
\end{align}
with matrix elements given by
\begin{align}
\frac{\partial \mathcal{E}_j^{(n)}}{\partial\Lambda_j^{(p)}}=
\begin{cases}
0 &p>n+1\\
1-\frac{d_j}{n+1} &p=n+1\\
\sum_{i\neq j} \frac{d_in!g^{n-p+1}}{p!(\epsilon_i-\epsilon_j)^{n-p+1}}+2\binom{n}{p}\Lambda_j^{(n-p)} &p<n\\
-1+\sum_{i\neq j}g \frac{d_i}{\epsilon_i-\epsilon_j}+2\Lambda_j^{(0)} &p=n\\
\end{cases}
\end{align}

\noindent and for $i\neq j$

\begin{align}
\frac{\partial \mathcal{E}_i^{(n)}}{\partial\Lambda_j^{(p)}}=
\begin{cases}
-d_j n! \frac{g^{n+1}}{(\epsilon_j-\epsilon_i)^{n+1}} &p=0\\ 
0 &p\neq 0.
\end{cases}
\end{align}
To have an idea of the structure of $S$ consider a $N=3$ system with $d_j=3~\forall j$, we would have
\begin{align*}
S=\begin{pmatrix}
\begin{pmatrix}
* & * & 0\\
* & * & *\\
* & * &* 
\end{pmatrix} & 
\begin{pmatrix}
* & 0 & 0 \\
* & 0 & 0 \\
* & 0 & 0 
\end{pmatrix} & 
\begin{pmatrix}
* & 0 & 0 \\
* & 0 & 0 \\
* & 0 & 0 
\end{pmatrix}\\
\begin{pmatrix}
* & 0 & 0 \\
* & 0 & 0 \\
* & 0 & 0 
\end{pmatrix} & 
\begin{pmatrix}
* & * & 0\\
* & * & * \\
* & * & * 
\end{pmatrix} & 
\begin{pmatrix}
* & 0 & 0 \\
* & 0 & 0 \\
* & 0 & 0 
\end{pmatrix}\\
\begin{pmatrix}
* & 0 & 0 \\
* & 0 & 0 \\
* & 0 & 0 
\end{pmatrix}& 
\begin{pmatrix}
* & 0 & 0 \\
* & 0 & 0 \\
* & 0 & 0 
\end{pmatrix} &
\begin{pmatrix}
* & * & 0 \\
* & * & * \\
* & * & * 
\end{pmatrix}\end{pmatrix},
\end{align*}
which does contain a lot of zeros. It should be obvious that for very large systems with high degeneracies this sparse structure of $S$ will play an important role with respect to computation time.

The Taylor expansion allows us to take larger steps by taking into account the derivatives of $\Lambda_{j}^{(n)}$ with respect to $g$. For a step $\delta g$ the first guess will thus be given by 
\begin{align}
\Lambda^{(n)}_j(g+\delta g)&\approx \sum_{k=0}^l \frac{(\delta g)^k}{k!}\frac{\partial^{k}\Lambda^{(n)}_j }{\partial g^{k}}
\nonumber\\&\equiv\sum_{k=0}^l \frac{(\delta g)^k}{k!}\Lambda^{(n,k)}_j 
\label{taylor}
\end{align}
where we note $\frac{\partial^{k}\Lambda^{(n)}_j }{\partial g^{k}}\equiv \Lambda_j^{(n,k)}$ the function $\Lambda_j$ differentiated $n$ times with respect to $\epsilon_j$ and $k$ times with respect to $g$.

The $g-$derivatives in (\ref{taylor}) can be found by recursively solving the linear system

\begin{align}
&A\vec{v}^{(k)}=\vec{R}^{(k)}&\\
&\vec{v}^{(k)}=
\left(\Lambda_1^{(0,k)},
\cdots,
\Lambda_1^{(d_1-1,k)},
\cdots,
\Lambda_{N}^{(0,k)},
\cdots,
\Lambda_{N}^{(d_{N}-1,k)}\right)&
\nonumber\\&
\vec{R}^{(k)}=\left(R^{(0,k)}_1,
\cdots,
R^{(d_1-1,k)}_1,
\cdots,
R^{(0,k)}_{N},
\cdots,
R^{(d_{N}-1,k)}_{N}
\right)\nonumber\\
\nonumber
\label{derivg}
\end{align}

\noindent and the elements of vector $\vec{R}^{(k)}$ are given by 


\begin{align}
R^{(n,k)}_j=&-\sum_{s=1}^{k-1}\binom{k}{s}\sum_{l=0}^n\binom{n}{l}\Lambda_j^{(l,s)}\Lambda_j^{(n-l,k-s)}
\nonumber\\&+\sum_{i\neq j}^{N}d_i n!\sum_{s=1}^k\binom{k}{s}s!
\nonumber\\&\times\left(\binom{n+1}{s}g^{n+1-s}\frac{\Lambda_i^{(0,k-s)}-\Lambda_j^{(0,k-s)}}{(\epsilon_i-\epsilon_j)^{n+1}}\right.
\nonumber\\&\left.-\sum_{l=1}^n\binom{l}{s}g^{l-s}\frac{\Lambda_j^{(n-l+1,k-s)}}{(n-l+1)!(\epsilon_i-\epsilon_j)^l}\right).
\end{align}

Since only the rhs of (\ref{derivg}) depends on $k$ we will only need to decompose matrix $A$ once (using the usual approaches for linear systems such as QR Decomposition, LU Decomposition or Inversion, ...) and solve the linear system for the new vector $\vec{R}^{(k)}$ up to a fixed number of derivatives. After the first guess is obtained a typical Newton method can be used to finally get an accurate solution.

\section{Root Extraction}
\label{rootex}

\subsection{General setup}

Having obtained solutions for the set of variables $\{\Lambda_i\}$ previously defined, it remains necessary to extract the actual rapidities from this set. In fact, Slavnov's determinant formulas\cite{slavnov} which give compact representations for scalar products and matrix elements, is yet only defined in terms of $\{\lambda_i\}$.  In simple terms, one needs to find the roots of the polynomial $P(z)$ which ultimately correspond to a given solution of the Bethe equations (\ref{bethe}).

While this constitutes a standard root-finding problem, the position of the roots in the complex plane can lead to numerical difficulties. In previous papers \cite{baode,buccheri} the monomial expansion $P(z) = \sum_{n=0}^M P_n(\{\lambda_i\})z^n$ was used. In this case, the coefficients are simply given in terms of the elementary symmetric polynomials of the set $\{\lambda_i\}$ which, in principle, can be found by solving a linear system of equations. For a non-degenerate model one would use $M$ equations given by

\begin{align}
P(\epsilon_j)\Lambda(\epsilon_j)&=P'(\epsilon_j)\nonumber\\
\Lambda(\epsilon_j)\sum_{m=0}^M\epsilon_j^m P_{M-m}&=\sum_{m=0}^M m \epsilon_j^{m-1} P_{M-m}.
\label{polylin}
\end{align}

However as shown by Wilkinson's numerical studies \cite{wilkinson}, expressing a polynomial in terms of its monomial expansion can give rise to serious numerical problems. In fact, the evaluation of the polynomial at a point $z$ far from 0 becomes very sensible to the finite numerical precision of the coefficients at large powers. At the end, the numerical error becomes rapidly sufficient to affect even the structure (real vs. complex conjugate pairs) of the roots. In order to circumvent that problem we use an alternative representation of the $P(z)$ polynomial by decomposing it onto the basis of Lagrange polynomials just like for polynomial interpolation. Picking any grid of $N_G=M+1$ distinct points $z_i$ and the corresponding values $P(z_i)$ one can exactly write

\bea
P(z) = \ell(z) \sum_{i=1}^{N_G} \frac{w_iP(z_i)}{z-z_i}\equiv \ell(z) \sum_{i=1}^{N_G} \frac{u_i}{z-z_i}
\label{barycentric}
\eea

\noindent where we defined $\displaystyle \ell(z) \equiv \prod_{i=1}^{M+1} (z-z_i)$, the barycentric weights $\displaystyle w_i = \frac{1}{\prod_{j=1,j\ne i}^{M+1} (z_i-z_j)}$ and $u_i \equiv w_iP(z_i)$. While $N_G=M+1$ points is a minimal requirement to represent a general polynomial of order $M$, here only $M$ points are in fact necessary since the $z^M$ coefficient is trivially 1. 

From the Riccati ODE (\ref{riccati}) it is simple to show that the polynomial obeys the following second order ODE:

\bea
P''(z) - F(z)P'(z)+G(z)P(z) = 0
\label{secondODE}
\eea

\noindent with 
\bea
F(z) &=& \frac{C}{g}+\frac{Bz}{g} + \sum_{j=1}^N \frac{d_j}{z-\epsilon_j}
\nonumber\\
G(z) &=&\frac{MB}{g} +\sum_{j=1}^N \frac{d_j \Lambda(\epsilon_j)}{z-\epsilon_j} 
\eea

Independently of the degeneracies $d_j$ it becomes possible to write a sufficient large system of linear equations whose solution fully specifies the polynomial $P(z)$. Indeed the first two derivatives of the barycentric representation (\ref{barycentric}) are easily shown to be given by

\bea
\frac{P(z)}{\ell(z)} &=&  \sum_{i=1}^{M+1} \frac{u_i}{z-z_i},
\nonumber\\
\frac{P'(z)}{\ell(z)} &=& \sum_{ i \ne j (=1)}^{M+1} \frac{u_i}{(z-z_i)(z-z_j)},
\nonumber\\
\frac{P''(z)}{\ell(z)} &=& \sum_{i\ne j \ne k (=1)}^{M+1} \frac{u_i}{(z-z_i)(z-z_j)(z-z_k)}.
\eea

Provided $z_i$ differs from every $\epsilon_j$ (so that $F(z_i),G(z_i)$ remain finite), Eq. (\ref{secondODE}) evaluated at $z = z_i$ leads to the linear (in the coefficients $u_j$) equation:

\bea
&&\sum_{ j \ne k (\ne i)} \frac{u_i}{(z_i-z_j)(z_i-z_k)} +2 \sum_{j \ne k (\ne i)}^{M+1} \frac{u_j}{(z_i-z_j)(z_i-z_k)}\nonumber\\ &&- F(z_i)\left(\sum_{ j  \ne i} \frac{u_i}{(z_i-z_j)} +\sum_{ j  \ne i} \frac{u_j}{(z_i-z_j)}  \right)
+G(z_i) u_i=0\nonumber\\
\eea

For a grid point $z_i = \epsilon_i$,  one can use the simpler $P'(\epsilon_i) = \Lambda(\epsilon_i) P(\epsilon_i)$ and find the equation:

\bea
 \sum_{  j (\ne i)}^{M+1} \frac{u_i}{(\epsilon_i-z_j)} + \sum_{  j (\ne i)}^{M+1} \frac{u_j}{(\epsilon_i-z_j)}
= \Lambda(\epsilon_i) u_i.
\eea

For any given grid $\{z_i\}$, one can therefore obtain the coefficients $\{u_i\}$ of the Lagrange basis representation simply by solving a set of linear equations. With this set of coefficients we have a representation of the polynomial from which we can extract its zeros using the standard Laguerre's method with deflation (i.e. dividing by $(x-\lambda)$ whenever a root $\lambda$ is found). In this representation deflation is a very simple task as well since it can be performed by shifting one grid point, say $z_1$, to the new found root $\lambda$. In doing so we simply set $u_1 \equiv w_1P(\lambda) =0$ since the polynomial has a root at that point. At the other points we have $u_j \to u_j \frac{z_j-z_1}{z_j-\lambda}$ due to the change in the barycentric weight $w_i$. We can then reduce the degree of the polynomial by one by simply removing the $z_1 \to \lambda$ point and repeat the procedure using the remaining $N_G-1$ grid points.

In this construction, we are free to choose the grid $\{z_i\}$ as we please and this fact can be exploited to ensure a numerically stable calculation. Indeed, this representation of the polynomial is heavily sensitive to numerical precision only when evaluating the polynomial at points $z$ which are far from any grid point $z_i$. Provided one can define a grid which is close enough to the actual roots $z_i(g) \approx \lambda_i(g)$, the impact of finite numerical precision can be controlled. In any case, the remaining error on the rapidities extracted from this procedure can always be corrected by refining the values using the original Bethe equations (\ref{bethe}). One should understand that an arbitrary set of $z_i$ cannot provide a numerically stable extraction of the rapidities. However, contrarily to any fixed polynomial basis (such as the monomial expansion for example), the Lagrange basis always makes it possible to choose an appropriate interpolation grid which leads to a stable result.

\subsection{Choosing the grid}

Whenever one is interested in studying the system for a wide variety of coupling strengths, the root extraction should then be performed at a number of intermediate points in the coupling strength scan. One could then simply define the grid points (at $g+\Delta g$) as the roots found at the previous point $z_i(g+\Delta g) = \lambda_i(g)$ which guarantees proximity of the grid to the actual solution and should always lead to numerically stable evaluation of the polynomial.

However, when studying a single given value of the coupling strength it is prohibitively time-consuming to perform the root extraction at intermediate points and one should instead exploit our knowledge of  the properties of any given solution to the Bethe equations to define an appropriate set of points $z_i(g)$ which mimics the positions of roots to be found.

At weak coupling, every rapidity is contained within the bandwidth of the energy levels $[\epsilon_1,\epsilon_N]$ with each roots corresponding to a given energy level. It is then a trivial task to define a grid which is close to the roots. In the particular cases treated here and shown in Figures (\ref{raps60}) and (\ref{raps128}) we choose two different sets of grid points. The first set, being the one which provides solutions to the BA at weak $g$. Here we use the linearized solution given by Equation (\ref{ans}) and substitute $g\rightarrow 0.1$, in this way we ensure proximity to the actual solution but at the same time avoid the risk of having to evaluate $\frac{P(z)}{\ell(z)}$ at $z\rightarrow z_j$. At some point in the computation, the actual solution approaches the grid and the evaluation of $\frac{P(z)}{\ell(z)}$ leads to a loss of stability. Since this occurs at a different point in $g$ depending on the state and the system size we simply use the original BA equations to determine how close the extracted roots are to the right solution. When a criterion on the precision is no longer met, we simply set the new grid to be equal to the rapidities calculated at the previous point keeping this grid for the remainder of the calculation.

When interested only in performing the root extraction at strong coupling one can define a adequate grid without having to find roots at intermediate values. In this strong coupling regime we know that a subset of $n_{div}$ rapidities will diverge while the remaining $M-n_{div}$ stay finite with real parts contained within the previously mentioned bandwidth (see Figures (\ref{raps60}) and (\ref{raps128}) for examples). We therefore need a grid defined by $M-n_{div}$ points $z_i$ within the bandwidth and $n_{div}$ points which diverge in order to follow the diverging roots.

 In \cite{faribault2} it was shown that  for non-degenerate spin-$\frac{1}{2}$ systems, provided $M \le N/2$ (\cite{notepv}), the configuration of roots at zero coupling is sufficient to determine the number $n_{div}$ of roots that will diverge in the strong coupling limit via a simple procedure. It simply relies on defining contiguous blocks of $\uparrow$ and $\downarrow$ spins. Starting from the highest energy level, the first $\uparrow$ block has size $P_1$, the second $P_2$ ..., while the contiguous blocks of $\downarrow$ separating them have sizes $H_1,H_2, ...$  With $N_b$ the total number of blocks, we have:\bea
n_{div} = [P_{N_b} + \alpha_{N_b-1}-\mathrm{Min}(P_{N_b} + \alpha_{N_b-1}, H_{N_b} )] 
\eea

\noindent with the $\alpha_i$ terms defined recursively as:
\bea
\alpha_0 &=& 0,
\nonumber\\
\alpha_i &=& [P_i + \alpha_{i-1} -\mathrm{Min}(P_i + \alpha_{i-1}, H_{i})] .
\eea

In the degenerate case, the procedure remains exactly the same. We simply need to consider the degeneracies to be slightly lifted and consider any rapidities $\lambda=\epsilon_i$ to flip the spin at the "bottom" of the corresponding set of degenerate energies as in Fig. \ref{blocks}'s example .

\begin{figure}[h]
\begin{equation}
\begin{array}{ccccccccc}
\circ\circ\circ\circ & \ \ &\bullet\bullet\circ\circ  & \ \ &\bullet\bullet\bullet\bullet & \ \ &\bullet\circ\circ & \ \ &\bullet  \\
\epsilon_1&\ \ & \epsilon_2 & \ \ &\epsilon_3 & \ \ &\epsilon_4& \ \ &\epsilon_5
\end{array}
\end{equation}
\caption{Graphical representation of the state $\{\lambda(g=0)\} = \{\epsilon_2,\epsilon_2,\epsilon_3,\epsilon_3,\epsilon_3,\epsilon_3,\epsilon_4,\epsilon_5\}$ for a configuration with degeneracies $\{d_1 = 4, d_2 = 4, d_3 = 4, d_4 = 3, d_5 = 1\}$. The black dots correspond to $\uparrow$ (spins flipped from the pseudo-vacuum). From the counting of blocks, this particular state would have 1 diverging rapidity in the $g\to \infty$ limit. }
\label{blocks}
\end{figure}

Remarkably, for a given $n_{div}$, independently of the set of $\epsilon_i$ and their degeneracies, we know \cite{altshuler} that in the $g\to \infty$ limit, these rapidities tend to $g L_i$ where $L_i $ are the $n_{div}$ roots of the Laguerre polynomial $L^{-1-\mathcal{N}-2n_{div}+2M}_{n_{div}}$. While one could numerically evaluate the exact location of these roots $L_i$, it is sufficient to simply define a set of points which covers the known support of these roots. Indeed,  the real and imaginary parts of the complete set  $\{L_i\}$ all fall within an easily defined bounded region of the complex plane\cite{petro} :
\begin{align}
-\mathcal{N}-2n_{div}+2M\le \Re (L_i) \le -\mathcal{N}+2M-2
 \nonumber\\
 |\Im (L_i)| \le2\sqrt{-n_{div}(-\mathcal{N}-2n_{div}+2M)}.
\end{align} 

One can then use $n_{div}$ points defined by real $z_i$ equally spaced within the interval $[\epsilon_1 + (-\mathcal{N}-2n_{div}+2M)g,\epsilon_1+(-\mathcal{N}+2M-2) g]$ combined with $M-n_{div}$ points located within the bandwidth. This choice proves sufficient to guarantee a numerically stable evaluation of the polynomial $P(z)$ in both regions where the zeros (and therefore rapidities) are to be found. One can then apply a refining procedure using the original $\lambda$-dependent Bethe equations (\ref{bethe}). 

For any given Hamiltonian and any particular state, the aforementioned ideas should suffice to build an adequate grid at strong coupling. In this regime, the splitting of rapidities into a diverging and finite block makes the choice of grid points central to the stability of the algorithm. In any case, it should also be possible to use the algorithm starting with any given grid and reuse the obtained solution to define a new grid. This would lead to an iterative process which, at first, evaluates the rapidities with a fairly large numerical error due to a poor choice of grid points. However, successive steps would see the error reduced  by a better and better choice of grid.


%

\section{Reduced BCS on a square lattice}
\label{bcs}

In order to demonstrate the capabilities of this method we now apply it in order to find a few eigenstates of a particular degenerate Gaudin model. The two dimensional Hubbard model can be written in Fourier space as
\begin{align}
H=&-2t\sum_{\vec{k},\sigma}(\cos k_x+\cos k_y)\hat{c}^\dagger_{\vec{k},\sigma}\hat{c}_{\vec{k},\sigma}\nonumber\\
&-\frac{U}{L^2}\sum_{\vec{k}_1,\vec{k}_2,\vec{q}}\hat{c}^\dagger_{\vec{k}_1,\uparrow}\hat{c}^\dagger_{\vec{k}_2,\downarrow}\hat{c}_{\vec{k}_2-\vec{q},\downarrow}\hat{c}_{\vec{k}_1+\vec{q},\uparrow}.
\label{hubbard}
\end{align}
If we restrict to $\vec{k}_1=-\vec{k}_2=\vec{k}$ we can write (\ref{hubbard}) in the form of a reduced BCS model,
\begin{align}
H=&-2t\sum_{\vec{k},\sigma}(\cos k_x+\cos k_y)\hat{c}^\dagger_{\vec{k},\sigma}\hat{c}_{\vec{k},\sigma}\nonumber\\
&-\frac{U}{L^2}\sum_{\vec{k},\vec{k}'}\hat{c}^\dagger_{\vec{k},\uparrow}\hat{c}^\dagger_{-\vec{k},\downarrow}\hat{c}_{-\vec{k}',\downarrow}\hat{c}_{\vec{k}',\uparrow},
\label{square}
\end{align}
which corresponds to the Richardson model\cite{rs-62} with interaction parameter $g=\frac{U}{L^2}$ and single particle energies $\epsilon_{\vec{k}}=-2t(\cos k_x+\cos k_y)$.

In the examples shown below we set $t=1$ and define $\epsilon_j$ for a set of points $(k_x,k_y)=\frac{2\pi}{L}(n_x,n_y)$ with $(n_x,n_y)=\{0,\dots,L\}$.

\begin{figure}[h]
\mbox{$M=60$~~~~$N=19$~~~~$\mathcal{N}=121$}\\
\includegraphics[width=0.23\textwidth]{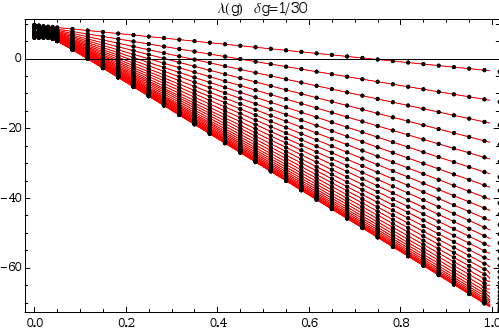}\includegraphics[width=0.23\textwidth]{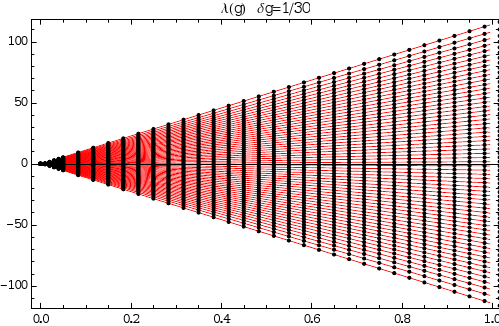}\\
\includegraphics[width=0.23\textwidth]{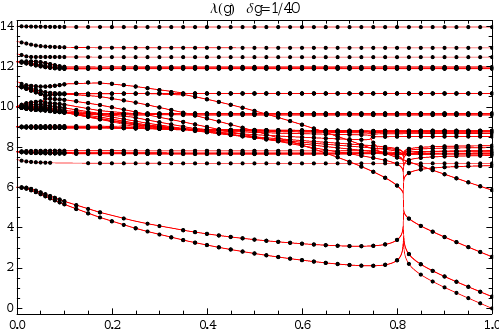}\includegraphics[width=0.23\textwidth]{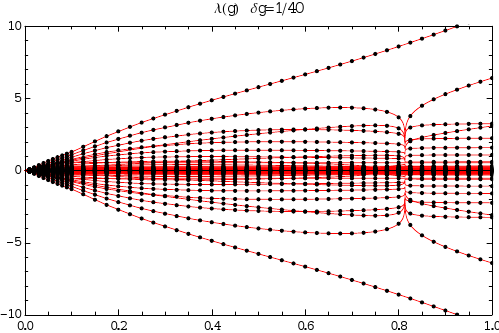}\\
\includegraphics[width=0.23\textwidth]{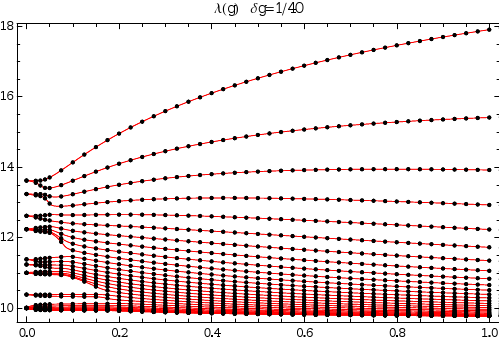}\includegraphics[width=0.23\textwidth]{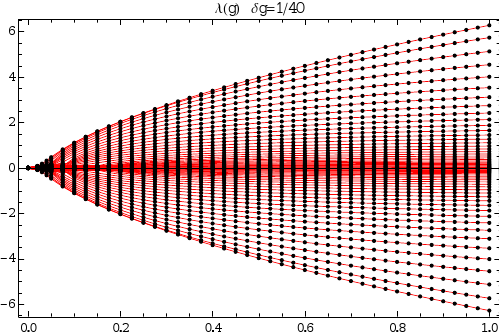}\\
\includegraphics[width=0.23\textwidth]{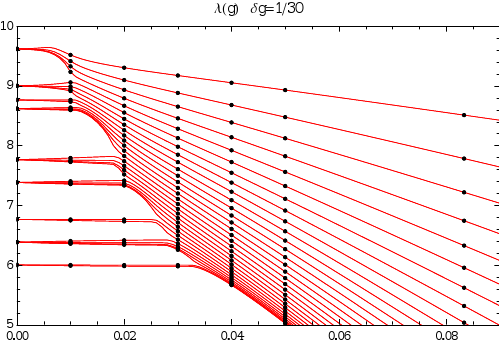}\includegraphics[width=0.23\textwidth]{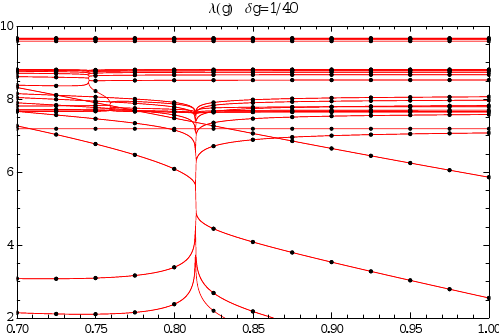}\\
\includegraphics[width=0.23\textwidth]{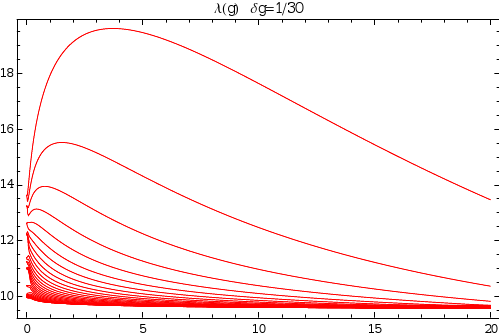}
\caption{On the top the real and the imaginary parts of the rapidities of the ground state with respect to $g$, the black dots represent the points at which the roots are actually computed. Step size is $\delta g=1/100$ from $0$ to $0.05$ and $\delta g=1/30$ from $0.05$ to $1$. The second and third figure from the top show the real and imaginary parts of a small excitation and a large excitation respectively, the step size is $\delta g=1/100$ from $0$ to $0.1$ and $\delta g=1/40$ from $0.1$ to $1$ for the small excitation and $\delta g=1/100$ from $0$ to $0.05$ and $\delta g=1/40$ from $0.05$ to $1$ for the large excitation. Below we show a focus around the critical region (where the rapidities meet/separate to become complex/real) for the ground state and the lowly excited state. At last we show the behavior of the highly excited state's rapidities on an extended region going from $g=0$ to $g=20$, step size is $\delta g=1/30$.}
\label{raps60}
\end{figure}

\begin{figure}[h]
\mbox{$M=128$~~~~$N=36$~~~~$\mathcal{N}=256$}\\
\includegraphics[width=0.23\textwidth]{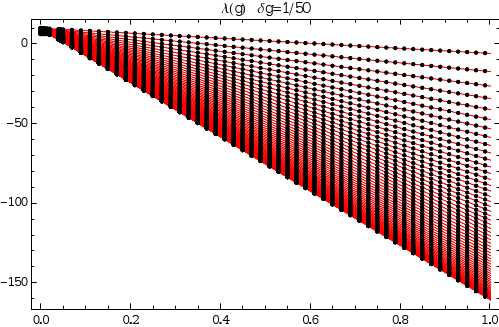}\includegraphics[width=0.23\textwidth]{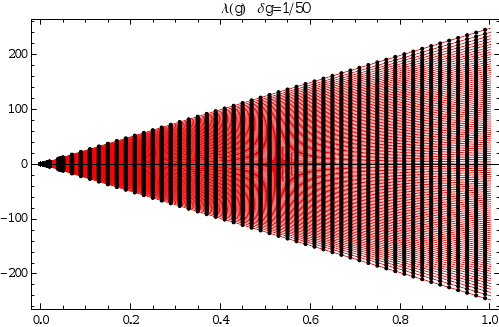}\\
\includegraphics[width=0.23\textwidth]{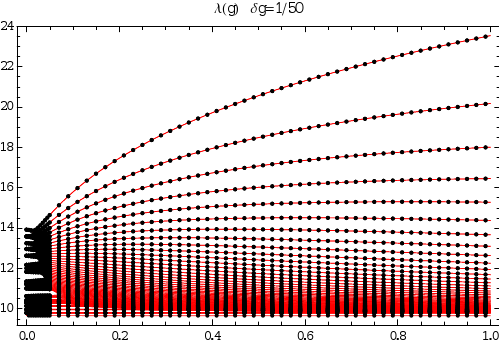}\includegraphics[width=0.23\textwidth]{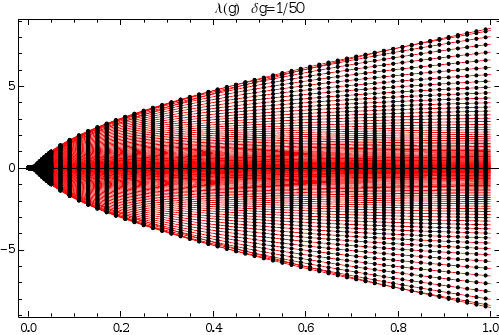}\\
\includegraphics[width=0.23\textwidth]{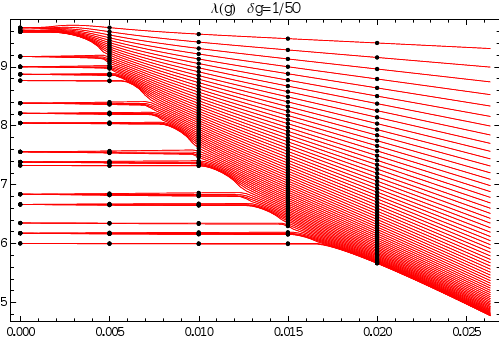}\includegraphics[width=0.23\textwidth]{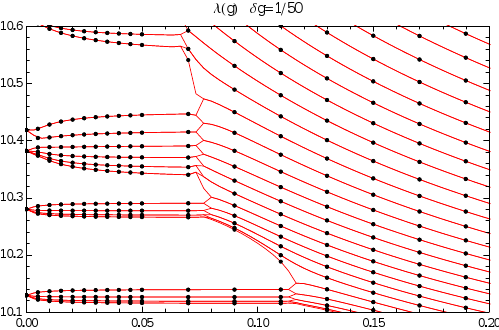}
\caption{Rapidities of the ground state and an excited state. Step size is $\delta g=1/200$ from $0$ to $0.05$ and $\delta g=1/50$ from $0.05$ to $1$.}
\label{raps128}
\end{figure}

Figures (\ref{raps60},\ref{raps128}) show the behavior of the rapidities for the degenerate energies from (\ref{square}) for the ground state and some excited states of two different sized systems ($L=10$ and $L=15$). In the $L=10$ case there are $N=19$ distinct values of $\epsilon_j$ with $4$ distinct degeneracies $d=\{1,4,8,20\}$,
in the $L=15$ case there are $N=36$ distinct values of $\epsilon_j$ with $2$ distinct degeneracies $d=\{4,8\}$.

As one can see, the steps in $g$ which can be taken while maintaining stability can be quite large compared to the rate of change of the rapidities with respect to $g$. Even in this degenerate case, this allows a rapid scan in the coupling constant opening the possibility of solving a large number of eigenstates in a reasonable amount of computation time.


One should note that the step size could easily be increased with increasing $g$ without affecting the stability in any way. In fact, the choice of the step size only depends on the behavior of the $\Lambda_j$ functions and can in principle be fined-tuned in a particular implementation (see \cite{baode} for more information).

\section{Conclusion}

In this work, we showed how one can exploit the ODE/BA correspondence to design an efficient numerical approach for finding solutions to the Bethe equations defining eigenstates of Gaudin models when degeneracies are present. It ultimately results in finding solutions to a set of quadratic equations expressed in terms of new variables. Moreover, it tuns out that in such an approach, degeneracies or equivalently (pseudo-)spins of magnitude larger than $\frac{1}{2}$ give rise to a sparse matrix structure making them actually simpler to treat than an equally large non-degenerate system. Examples for the reduced BCS Hamiltonian on a  square lattice were shown, demonstrating the efficiency of the method.

We also introduced a new numerical technique to extract actual rapidities from our intermediate variables. Being based on the Lagrange polynomials barycentric interpolation it offers much better numerical stability than previous suggestions based on a monomial expansion. All in all the techniques discussed here offer a remarkably fast and efficient algorithm for systematically finding solution to the Bethe equations.

The numerical stability and computation speed obtained via this method opens the possibility of studying non-equilibrium dynamics of such systems. Since it allows one to compute a large number of eigenstates, when possible, it could be combined with some truncation scheme allowing one to reduce the effective Hilbert space (see \cite{faribault} for such an example) . In more general settings, it can also allow one to use a simple Monte Carlo approach which would sample a large part of the Hilbert space. 

\section*{Acknowledgements}
O. E. A. and V. G. are supported by Swiss National Science Foundation. A. F. is supported by the German Research Foundation (DFG).

\begin{appendix}
\section{Derivation of $\mathcal{E}^{(n)}$}
\label{derivation}
Deriving the equation for the $n^{th}$ derivative of
\begin{align}
\mathcal{E}(z)=\Lambda'(z)+\Lambda^2(z)-\sum_{\alpha=1}^N\frac{2F(\lambda_\alpha)}{z-\lambda_\alpha}=0.
\label{lambdaeqapp}
\end{align}
is mostly a technical task, which this appendix adresses by showing how to obtain the derivatives for the most general $F(\lambda_k)$. We are interested in
\begin{align}
F(\lambda_k)&=-\sum_{i=1}^N\frac{A_i}{(\epsilon_i-\lambda_k)}+\frac{B}{2g}\lambda_k+\frac{C}{2g}\nonumber\\
&\equiv -\sum_{i=1}^N\frac{A_i}{(\epsilon_i-\lambda_k)}+f(\lambda_k),
\label{F2}
\end{align}
were we defined $f(\lambda_k)=\frac{1}{2g}(B\lambda_k+C)$.

Inserting (\ref{F2}) in Eq. (\ref{lambdaeqapp}) we get 
\begin{align}
&\Lambda'(z)+\Lambda^2(z)-\sum_{\alpha=1}^M \frac{2f(\lambda_\alpha)}{z-\lambda_\alpha}
\nonumber\\&+\sum_{\alpha=1}^M\sum_{i=1}^N\frac{2 A_i}{(z-\lambda_\alpha)(\epsilon_i-\lambda_\alpha)}=0.
\label{lamF}
\end{align} 

The first three terms of (\ref{lamF}) are easily derived with the help of Leibniz relation
\begin{align}
(h g)^{(n)}=\sum_{k=0}^n\binom{n}{k} h^{(k)}g^{(n-k)}.
\label{leib}
\end{align} 

We then focus here in the derivation of the last term. Defining $\Lambda(z)^{(n)}$ as the $n^{th}$ derivative of the function $\Lambda(z)=\sum_{k=1}^M\frac{1}{z-\lambda_k}$ with respect to $z$, we have 
\begin{align}
\Lambda(z)^{(n)}=(-1)^n n!\sum_{k=1}^M\frac{1}{(z-\lambda_k)^{n+1}}.
\label{lambdader}
\end{align}
The $n^{th}$ derivative is then given by
\begin{align}
&\left(\sum_{\alpha=1}^M\sum_{i=1}^N\frac{d_i}{(z-\lambda_\alpha)(\epsilon_i-\lambda_\alpha)}\right)^{(n)}
\nonumber\\&=(-1)^n n!\sum_{\alpha=1}^M\sum_{i=1}^N\frac{d_i}{(\epsilon_i-\lambda_\alpha)(z-\lambda_\alpha)^{n+1}}\equiv(*).
\label{partb}
\end{align}
To write (\ref{partb}) in terms of $\Lambda(\epsilon_j)$ one has to take the limit $z\rightarrow\epsilon_j$. The term in the sum can then be expanded in the form
\begin{align}
&\frac{1}{(\epsilon_i-\lambda_\alpha)(\epsilon_j-\lambda_\alpha)^{n+1}}=
\nonumber\\&\left(\frac{1}{\epsilon_i-\lambda_\alpha}-\frac{1}{\epsilon_j-\lambda_\alpha}\right)\frac{1}{(\epsilon_j-\epsilon_i)^{n+1}}
\nonumber\\&-\sum_{k=1}^n\frac{1}{(\epsilon_j-\epsilon_i)^k(\epsilon_j-\lambda_\alpha)^{n-k+2}}.
\label{expa}
\end{align}
Inserting (\ref{expa}) in (\ref{partb}) we find

\begin{align}
(*)=&
\nonumber\\\sum_{\alpha=1}^M\sum_{i=1}^N&(-1)^n
n!\left[\left(\frac{1}{\epsilon_i-\lambda_\alpha}-\frac{1}{\epsilon_j-\lambda_\alpha}\right)\frac{1}{(\epsilon_j-\epsilon_i)^{n+1}}\right.
\nonumber\\&\left.-\sum_{k=1}^n\frac{1}{(\epsilon_j-\epsilon_i)^k(\epsilon_j-\lambda_\alpha)^{n-k+2}}\right]
\nonumber\\=&-\sum_{i=1}^N d_i n!\left(\frac{\Lambda(\epsilon_i)-\Lambda(\epsilon_j)}{(\epsilon_i-\epsilon_j)^{n+1}}\right.
\nonumber\\&\left.-\sum_{k=1}^n\frac{\Lambda(\epsilon_j)^{(n-k+1)}}{(\epsilon_i-\epsilon_j)^k}\frac{1}{(n-k+1)!}\right)\equiv(**)
\label{summ}
\end{align}

\noindent where the sum over $\alpha$ was substituted with the functions $\Lambda(\epsilon_j)$ and their derivatives using Eq. (\ref{lambdader}).

One only needs now to get rid of the divergence given by the term $i=j$ in the sum over $i=1,\dots,N$. Since we have $\epsilon_i-\epsilon_j\rightarrow\delta\rightarrow 0$ the fraction $\frac{\Lambda(\epsilon_i)-\Lambda(\epsilon_j)}{(\epsilon_i-\epsilon_j)^{n+1}}$ can be written in terms of a derivative of $\Lambda$. Therefore, writing $\Lambda(\epsilon_i+\delta)$ in its Taylor form,
\begin{align}
\Lambda(\epsilon_i+\delta)\approx&\sum_{k=0}^{n+1}\frac{\delta^k}{k!}\Lambda(\epsilon_j)^{(k)}=\Lambda(\epsilon_j)+\frac{\delta^{n+1}}{(n+1)!}\Lambda(\epsilon_j)^{(n+1)}
\nonumber\\&+\sum_{k=1}^{n}\frac{\delta^k}{k!}\Lambda(\epsilon_j)^{(k)}
\end{align}
one finds, for the term $i\rightarrow j$ of $(**)$
\begin{align}
(**)|_{i\rightarrow j}=&-d_jn!\left(\frac{\Lambda(\epsilon_j+\delta)-\Lambda(\epsilon_j)}{\delta^{n+1}}\right.
\nonumber\\&\left.-\sum_{k=1}^n\frac{\Lambda(\epsilon_j)^{(n-k+1)}}{\delta^k}\frac{1}{(n-k+1)!}\right)
\nonumber\\=&-d_j n!\left(\frac{1}{(n+1)!}\Lambda(\epsilon_j)^{(n+1)}\right.
\nonumber\\&\left.+\sum_{k=1}^{n}\frac{\delta^{k-n-1}}{k!}\Lambda(\epsilon_j)^{(k)}\right.
\nonumber\\&\left.-\sum_{k=1}^n\frac{\delta^{-k}}{(n-k+1)!}\Lambda(\epsilon_j)^{(n-k+1)}\right)
\nonumber\\=&-\frac{d_j}{n+1}\Lambda(\epsilon_j)^{(n+1)}
\end{align}
where the second and the third term cancel themselves since they correspond to the same sum with reversed indices $k=1,\dots,n\rightarrow \hat{k}=n,\dots,1$ and $\hat{k}\rightarrow n-k+1$.

The third term of (\ref{lamF}) is finally given by
\begin{align}
(*)=&-\sum_{i\neq j}^Nd_in!\left(\frac{\Lambda(\epsilon_i)-\Lambda(\epsilon_j)}{(\epsilon_i-\epsilon_j)^{n+1}}\right.
\nonumber\\&\left.-\sum_{k=1}^n\frac{\Lambda(\epsilon_j)^{(n-k+1)}}{(\epsilon_i-\epsilon_j)^k}\frac{1}{(n-k+1)!}\right)
\nonumber\\&-\frac{d_j}{n+1}\Lambda(\epsilon_j)^{(n+1)}.
\end{align}

Thus, using Leibniz equation (\ref{leib}) and noting that $f(\lambda_\alpha)^{(n)}=0~\forall~n>0$ the $n^{th}$ derivative of (\ref{lamF}) in the limit $z\rightarrow\epsilon_j$ reads
\begin{align}
\mathcal{E}(\epsilon_j)^{(n)}=&(1-\frac{d_j}{n+1})\Lambda(\epsilon_j)^{(n+1)}
\nonumber\\&+\sum_{k=0}^n\binom{n}{k}\Lambda(\epsilon_j)^{(k)}\Lambda(\epsilon_j)^{(n-k)}
\nonumber\\&-\frac{1}{g}\left(Bn\epsilon_j\Lambda(\epsilon_j)^{(n-1)}+C\Lambda(\epsilon_j)^{(n)}\right)
\nonumber\\&-\sum_{i\neq j}^N d_i n!\left(\frac{\Lambda(\epsilon_i)-\Lambda(\epsilon_j)}{(\epsilon_i-\epsilon_j)^{n+1}}\right.
\nonumber\\&\left.-\sum_{k=1}^n\frac{\Lambda(\epsilon_j)^{(n-k+1)}}{(\epsilon_i-\epsilon_j)^k}\frac{1}{(n-k+1)!}\right).
\label{matej}
\end{align}

\end{appendix}

\end{document}